\documentclass[9pt,conference]{IEEEtran}

\usepackage{cite}
\usepackage{amsmath,amssymb,amsfonts}
\usepackage{graphicx}
\usepackage{textcomp}
\usepackage{xcolor}
\usepackage{amsmath,graphicx,booktabs}
\usepackage{amssymb}
\usepackage{hyperref}
\usepackage{microtype}
\def\BibTeX{{\rm B\kern-.05em{\sc i\kern-.025em b}\kern-.08em
    T\kern-.1667em\lower.7ex\hbox{E}\kern-.125emX}}
\newcommand{\audiosym}{{\tt <|audio|>}}
\renewenvironment{quote}
  {\list{}{\leftmargin=1em \rightmargin=0.1em}\item\relax\small\ttfamily}
  {\endlist}

\begin{document}

\title{Granite-speech: open-source speech-aware LLMs with strong English ASR capabilities}



\author{\IEEEauthorblockN{George Saon\IEEEauthorrefmark{2}, Avihu Dekel\IEEEauthorrefmark{2}, Alexander Brooks\IEEEauthorrefmark{2}, Tohru Nagano\IEEEauthorrefmark{2},}
\IEEEauthorblockN{Abraham Daniels, Aharon Satt, Ashish Mittal, Brian Kingsbury, David Haws, Edmilson Morais,}
\IEEEauthorblockN{Gakuto Kurata, Hagai Aronowitz, Ibrahim Ibrahim, Jeff Kuo, Kate Soule, Luis Lastras, Masayuki Suzuki,}
\IEEEauthorblockN{Ron Hoory, Samuel Thomas, Sashi Novitasari, Takashi Fukuda, Vishal Sunder, Xiaodong Cui, Zvi Kons}
\vspace{0.2cm}
\IEEEauthorblockN{IBM Research}
\IEEEauthorblockN{\IEEEauthorrefmark{2}Core contributors}
}

\maketitle

\begin{abstract}
Granite-speech LLMs are compact and efficient speech language models specifically designed for English ASR and automatic speech translation (AST). The models were trained by modality aligning the 2B and 8B parameter variants of granite-3.3-instruct to speech on publicly available open-source corpora containing audio inputs and text targets consisting of either human transcripts for ASR or automatically generated translations for AST. Comprehensive benchmarking shows that on English ASR, which was our primary focus, they outperform several competitors' models that were trained on orders of magnitude more proprietary data, and they keep pace on English-to-X AST for major European languages, Japanese, and Chinese. The speech-specific components are: a conformer acoustic encoder using block attention and self-conditioning trained with connectionist temporal classification, a windowed query-transformer speech modality adapter used to do temporal downsampling of the acoustic embeddings and map them to the LLM text embedding space, and LoRA adapters to further fine-tune the text LLM. Granite-speech-3.3 operates in two modes: in speech mode, it performs ASR and AST by activating the encoder, projector, and LoRA adapters; in text mode, it calls the underlying granite-3.3-instruct model directly (without LoRA), essentially preserving all the text LLM capabilities and safety. Both models are freely available on HuggingFace (\url{https://huggingface.co/ibm-granite/granite-speech-3.3-2b} and \url{https://huggingface.co/ibm-granite/granite-speech-3.3-8b}) and can be used for both research and commercial purposes under a permissive Apache 2.0 license. 

\end{abstract}


\section{Introduction}
The landscape of speech (or spoken) language models (SLMs) is rapidly evolving. SLMs can be broadly classified into two categories: those that train on interleaved acoustic and text tokens and model the joint distribution of text and speech directly such as~\cite{team2023gemini,xie2024mini} and speech-aware LLMs, a terminology borrowed from~\cite{arora2025landscape}, that use an acoustic encoder and text instructions to perform a specific task on the audio content, the main examples being~\cite{grattafiori2024llama,abouelenin2025phi,ma2024embarrassingly,tangsalmonn,chu2024qwen2}. Both approaches have benefits and drawbacks. The first category, also known as early fusion models, seamlessly integrates audio and text early during the training, resulting in models that are fluent in both modalities. This comes at the expense of the model having fewer capabilities with audio prompts compared to the text-only model with corresponding text instructions, simply because instruction tuning and preference alignment from text are much more extensive. Such models are also inherently not as safe because there could be successful attacks with audio prompts that do not work with text prompts, as shown in~\cite{costa2024mutox}. In contrast, speech-aware LLMs require a delicate modality-alignment step that downsamples the acoustic embeddings coming out of the encoder to a rate comparable to the text embeddings and maps them to a space that is interpretable by the text LLM. This has the advantage that the text LLM can remain largely intact as exemplified in~\cite{grattafiori2024llama,ma2024embarrassingly,fan2024alignformer} or undergo minimal LoRA fine-tuning, thus preserving more text capabilities and safety/guardrails. The drawback is that mid-fusion models are not as fluent in both modalities and are limited to outputting text only, requiring an external text-to-speech module for speech generation.

Our proposed speech-aware LLMs operate in two steps. In step one, English speech input is transcribed or, optionally, translated into a different language. The output of step one can be fed, in step two, as a text prompt, potentially with a longer dialog context, to the underlying Granite text LLM to generate a response. Importantly, the text LLM is {\em shared} between the two steps, requiring only one instance in memory. 
During inference, we support two modes of operation: a text-only mode and a speech mode. The speech mode is automatically activated when a prompt includes both an \audiosym\ token and a corresponding audio file; otherwise, the model defaults to text-only mode. This setup allows users to toggle modes seamlessly by simply including or omitting the audio input.
In speech mode, the model runs with the acoustic encoder, speech modality adapter, and enables the LoRA adapter. In text-only mode, the text LLM is used with LoRA adapters disabled. 
By construction, this two-pass architecture ensures that all text capabilities such as retrieval-augmented generation, function-calling, and safety are preserved, at the cost of requiring two LLM generation calls.
This disadvantage can be mitigated through careful orchestration of the calls, for example, by caching previous key-value computations in the LLM attention layers. Compared to a cascaded dedicated ASR+LLM approach, our proposed model has the advantage of a stronger ASR component by leveraging the power of the LLM. We position the granite-speech-3.3 models as LLMs with good ASR/AST capabilities rather than standalone ASR/AST models (although they can certainly be used this way).

The paper is organized as follows. In Section~\ref{sec:system}, we discuss the overall system architecture, training data, encoder architecture, training and ASR experiments, architecture of the speech modality adapter, training and ASR experiments, speech translation experiments, and safety considerations. Section~\ref{sec:conclusion} summarizes our findings and suggests areas for future improvements.

\section{System description and experimental results}
\label{sec:system}

\subsection{Overall system architecture}
\begin{figure}[htb]
    \centering
    \includegraphics[width=0.98\columnwidth]{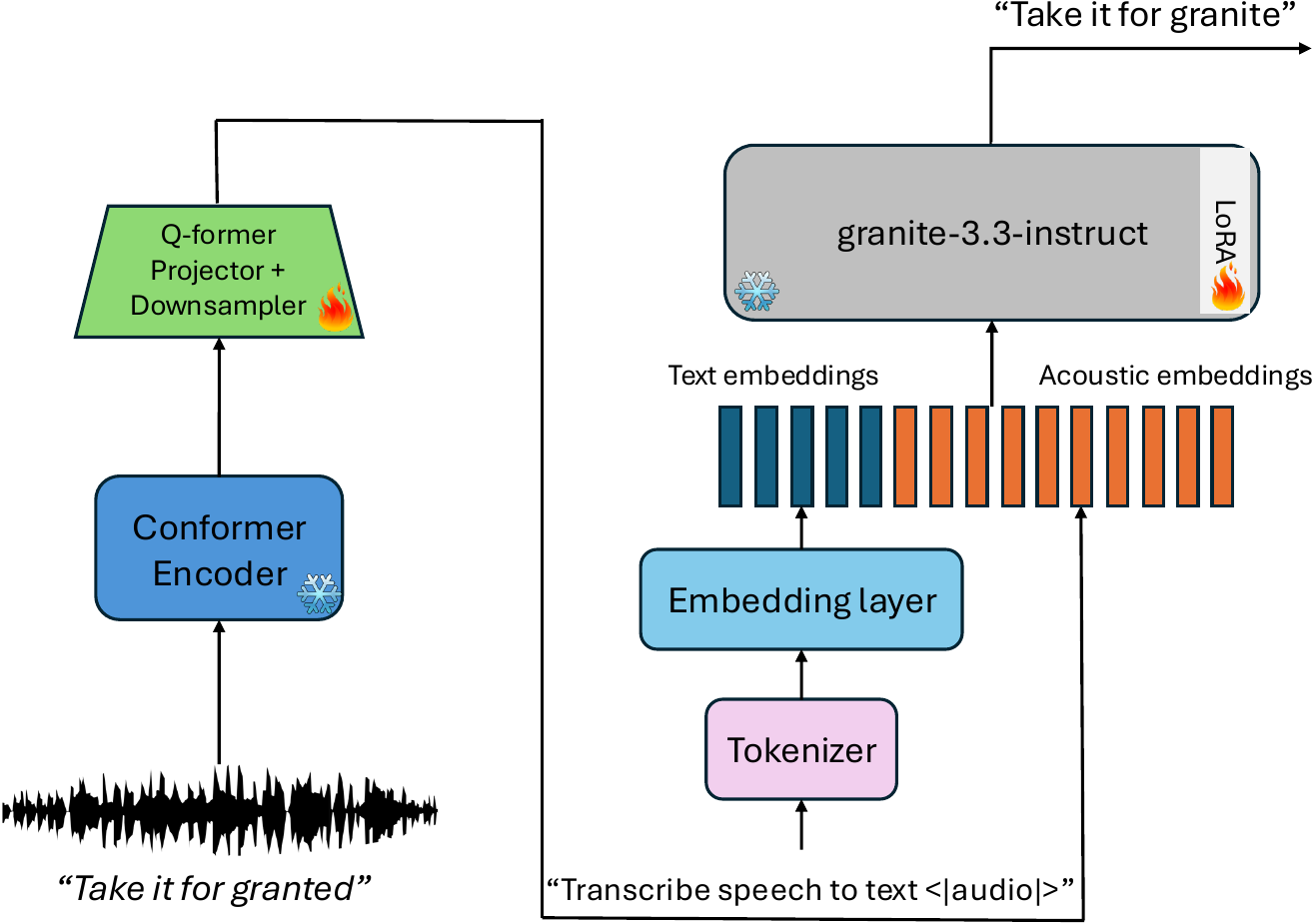}
    \caption{Overall system architecture.}
    \label{fig:architecture}
\end{figure}

Granite speech comprises the following components:
\begin{itemize}
\item Acoustic encoder used to convert the speech signal into a higher-level representation suitable for ASR and AST (described in subsection~\ref{ssec:encoder})
\item Speech modality adapter used to temporally downsample the acoustic embeddings and map them to a space that is interpretable by the text LLM (subsection~\ref{ssec:projector})
\item Granite text LLM used as the decoder component for the overall ASR/AST system or in isolation depending on whether the prompt contains audio or not
\item LoRA adapters applied to the query and value projection matrices in the attention blocks of the LLM layers used to fine-tune the LLM to the characteristics of the acoustic embeddings coming out of the modality adapter 
\end{itemize}
These modules are also illustrated in Figure~\ref{fig:architecture} where we indicate which modules are trainable/frozen during the joint projector/LLM training phase.

\subsection{Training data}
Our models are trained on major publicly available English ASR datasets as well as synthetic translations from CommonVoice English to French, Spanish, German, Italian, Portuguese, Japanese, and Chinese to support the speech translation task. In principle, this makes all of the reported experimental results reproducible by the research community. Depending on the experimental setup and the license requirements for the released models, training either included or excluded corpora with noncommercial licenses. Concretely, our models were trained on subsets of the following corpora: Multilingual LibriSpeech English~\cite{pratap2020mls}, Gigaspeech~\cite{chen2021gigaspeech}, CommonVoice 17.0~\cite{ardila2019common}, LibriSpeech~\cite{panayotov2015librispeech}, Voxpopuli~\cite{wang2021voxpopuli}, AMI~\cite{kraaij2005ami}, YODAS~\cite{li2023yodas}, SPGI Speech~\cite{o2021spgispeech}, Switchboard~\cite{godfrey1992switchboard}, CallHome, Fisher~\cite{cieri2004fisher}, Voicemail~\cite{padmanabhan2002automatic} and TED LIUM~\cite{rousseau2012ted}. The amount of audio data and the type of material for each corpus is summarized in Table~\ref{corpora}. 

\begin{table}[htb]
\begin{center}
\begin{tabular}{|l|l|r|} \hline
Corpus name & Material & Nb. of hours\\ \hline
MLS English & Audiobooks & 44000\\ \hline
GigaSpeech$^*$ & YouTube videos+podcasts & 10000\\ \hline
YODAS & YouTube videos & 10000\\ \hline
SPGI$^*$ & Earnings reports & 5000\\ \hline
CommonVoice 17 & User uploaded audio & 2600\\ \hline
Fisher & Telephone conversations & 2000\\ \hline
Librispeech & Audiobooks & 960\\ \hline
VoxPopuli & European parliamentary speeches & 500\\ \hline
Switchboard & Telephone conversations & 260\\ \hline
TED LIUM$^*$ & TED talks & 200\\ \hline
AMI & Meetings recordings & 100\\ \hline
Voicemail & Voicemail messages & 80\\ \hline
CallHome & Telephone conversations & 18\\ \hline
\end{tabular}
\end{center}
\caption{\label{corpora} Type of material and amount of audio data for each training corpus ($^*$ denotes corpora with noncommercial license that were excluded from the training of the final Apache 2.0 models).}
\end{table}

The modality and LoRA adapters leveraged synthetically generated speech translation training data, which was generated by translating the English section of CommonVoice 17 to major European languages as well as Japanese and Chinese. The choice of translation models as well as the filtering procedure are described in the AST subsection~\ref{ssec:ast}.

\subsection{Encoder architecture and training}
\label{ssec:encoder}

The speech encoder consists of a stack of conformer blocks~\cite{gulati2020conformer} trained with Connectionist Temporal Classification (CTC) on character-level targets. The exact configuration of the encoder is shown in Table~\ref{ctc-arch}. We use block attention in every conformer self-attention layer with a block size of 4 seconds similar to~\cite{zhang2023google}. Furthermore, the encoder is trained with self-conditioned CTC~\cite{nozaki2021relaxing} from the middle layer with a CTC loss weight of 0.2 for the intermediate layer and 0.8 for the final layer. 

\begin{table}[htb]
\begin{center}
\begin{tabular}{|l|c|} \hline
Configuration parameter & Value\\ \hline
Input dimension & 160 (80 logmels x 2) \\ \hline
Nb. of layers   & 10                   \\ \hline
Hidden dimension & 1024                \\ \hline
Nb. of attention heads & 8             \\ \hline
Attention head size    & 128           \\ \hline
Convolution kernel size & 15           \\ \hline
Output dimension        & 42           \\ \hline
\end{tabular}
\end{center}
\caption{\label{ctc-arch} Architecture details for the CTC speech encoder.}
\end{table}

The encoder is trained on 80-dimensional logmel features extracted every 10ms from 16kHz audio recordings. 
To reduce the number of tokens processed by the conformer, we perform temporal subsampling by a factor of 2 by stacking every two consecutive frames into a one vector. This yields 50 feature vectors per second, each 160-dimensional, which are then linearly projected to the input of the first conformer block.
Every audio sample is perturbed during training by adding various noises with probability (w.p.) 0.25 with an SNR in the range of $-5\ldots 20$ and performing SpecAugment~\cite{park19} w.p. 0.9 with time and frequency masking. The acoustic encoder is trained for 20 epochs (1.5M updates) with AdamW SGD with a batch size of 256 utterances using a triangular learning rate schedule that ramps up from 5e-5 to 5e-4 over the first 6 epochs and decays to 5e-6 over the next 14 epochs. At the beginning of every epoch, the utterances are randomly shuffled, divided into a fixed number of parts (typically 200), and sorted within each part by increasing acoustic sequence lengths. Batches are formed sequentially from the shortest to the longest utterances for the first part, then again from the shortest to the longest utterances for the second part, and so on until all parts are processed.

In Table~\ref{tok}, we compare the effect of the output tokenization on the performance of the CTC encoders in isolation with greedy decoding and also after joint LLM training for: characters (42 outputs), BERT uncased (32000 BPE units) and Granite tokenization (49000 BPE units). We selected character tokenization due to the better performance after joint LLM training where granite-3.1-8b-base was used as the base LLM.

\begin{table}[htb]
\begin{center}
\begin{tabular}{|l|c|c|c|c|c|c|c|c|c|}\hline
Tokenizer   &  CV  & GS & MLS & LSc & LSo & SPGI & AMIi & AMIs & Vox\\ \hline
Characters    & 14.5 & 12.5 & 6.7  & 1.7   & 4.0   & 4.6  & 11.9    & 31.0    & 8.1\\ 
+LLM     & 9.5  & 10.4 & 4.8  & 1.4   & 3.0   & 2.1  & 10.0    & 27.6    & 6.3\\ \hline
BERT     & 14.4 & 11.1       & 7.2  & 2.0   & 4.7   & 3.6  & 12.1    & 31.5    & 7.9 \\ \hline
Granite  & 14.5 & 11.2       & 7.1  & 2.2   & 4.8   & 3.1  & 10.9    & 28.9    & 7.7 \\ 
+LLM     & 9.8  & 10.2 & 4.9 & 1.4  & 3.2   & 2.0  & 10.3  & 28.5    & 6.1\\ \hline
\end{tabular}
\end{center}
\caption{\label{tok} Influence of output tokenization on word error rate for CTC speech encoders (with greedy decoding and after joint LLM training).}
\end{table}

The training data for the previous CTC encoders included corpora that had restrictive licenses for commercial use such as GigaSpeech, SPGI and TED\_LIUM. We retrained the CTC encoders by excluding those corpora and adding 10k hours of YODAS English data obtained by comparing user-uploaded transcripts with transcripts produced by Whisper medium. The recognition performance of two models with 10 layers, 275M parameters and 16 layers, 430M parameters is shown in Table~\ref{ctc-wer}. For chronological reasons, we used the smaller 10-layer encoder for all the following joint LLM experiments. 

\begin{table}[htb]
\begin{center}
\begin{tabular}{|l|c|c|c|c|c|c|c|c|c|}\hline
Encoder   &  CV  & GS & MLS & LSc & LSo & SPGI & AMIi & AMIs & Vox\\ \hline
10 layers &  13.3 &12.2 &6.7 &1.9 & 4.2 &  4.5 & 11.3 & 29.0 & 8.0\\ \hline
16 layers &  11.2 &11.7 &6.3 &2.2 & 4.2 &  4.4 & 10.9 & 28.2 & 7.8\\ \hline
\end{tabular}
\end{center}
\caption{\label{ctc-wer}ASR performance of CTC encoders with different number of layers trained only on corpora having Apache 2.0 compatible licenses (greedy decoding).}
\end{table}

\subsection{Task-specific prompt construction}
During both training and inference, we use the Granite chat formatting syntax, which consists of three turns: (1) a system prompt, (2) a user query, and (3) the model response. 
We adopt a fixed system prompt, that was used to train the Granite-3.3-instruct:
\begin{quote}
Knowledge Cutoff Date: April 2024. Today's Date: DATE. 
You are Granite, developed by IBM. You are a helpful AI assistant
\end{quote}

The user query can correspond to either a transcription (ASR) or a translation (AST). Each training example is labeled with a task tag, which determines how the prompt will be selected. 
Prompts contain a special \audiosym\ token, which gets replaced with the projected embeddings of the input audio. 
Given an ASR example, we randomly select a prompt from a set of 24 variations, e.g.: 
\begin{quote}
Listen to the speech and write down its content \audiosym.
\end{quote}
An AST example is randomly assigned into one of two task types: (a) direct speech translation, where the model generates the translation directly -- for example: 
\begin{quote}
\audiosym\ translate the speech to Spanish.
\end{quote}
or (b) chain-of-thought (CoT) speech translation, where the model first transcribes the text and then translates it, marking each part with explicit tags [Transcription] and [Translation]. This step-by-step approach has been shown to improve performance in prior work~\cite{abouelenin2025phi}. An example CoT-AST prompt:
\begin{quote}
\audiosym\ Can you transcribe the speech, and then translate it to Spanish?
\end{quote}
During training, each AST example is assigned to the CoT-AST variant with a probability of $p = 0.3$. 
Prompts are randomly selected from a pool of 24 AST prompts and 8 CoT-AST prompts.
The final turn in the chat sequence is the model response, which contains the expected output: a transcription for ASR, a translation for AST, and both transcription and translation for CoT-AST. 

After constructing the final textual input, we tokenize and embed it using the Granite tokenizer and text embedding table.
We then replace the \audiosym\ special token with the projected embeddings of the audio.
This combined representation is then passed to the LLM for generating the corresponding output.
To enable the LLM to consume a compact and informative representation of the input audio, we next describe the architecture of our modality adapter.

\subsection{Speech modality adapter architecture and training}
\label{ssec:projector}
\begin{figure}[htb]
    \centering
    \includegraphics[width=0.98\linewidth]{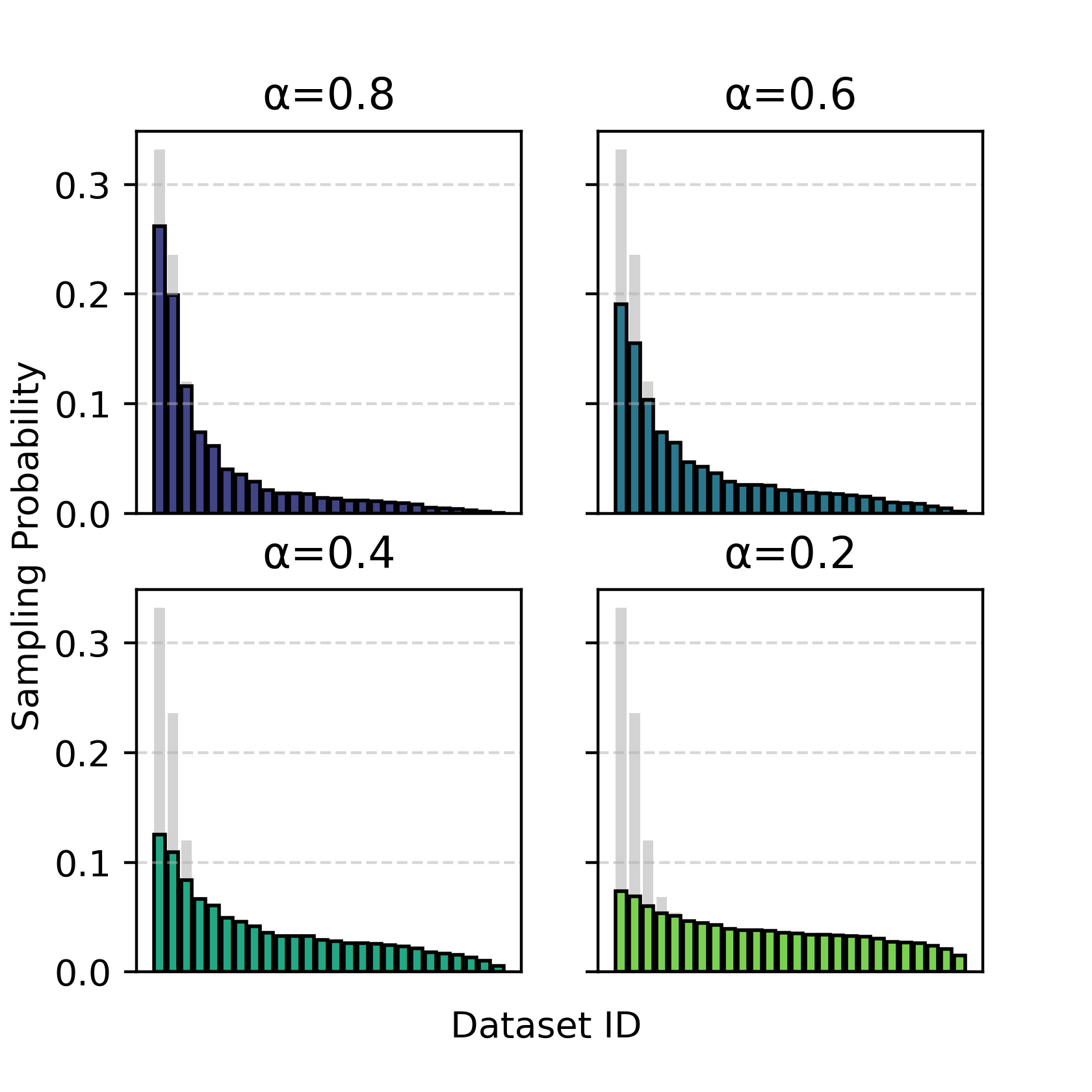}
    \caption{Visualizing how different $\alpha$ values in the balanced sampler affect the dataset distribution. Using lower $\alpha$ values yields a more balanced distribution.}
    \label{fig:balanced_sampler}
\end{figure}

\begin{figure}[htb]
    \centering
    \includegraphics[width=0.98\linewidth]{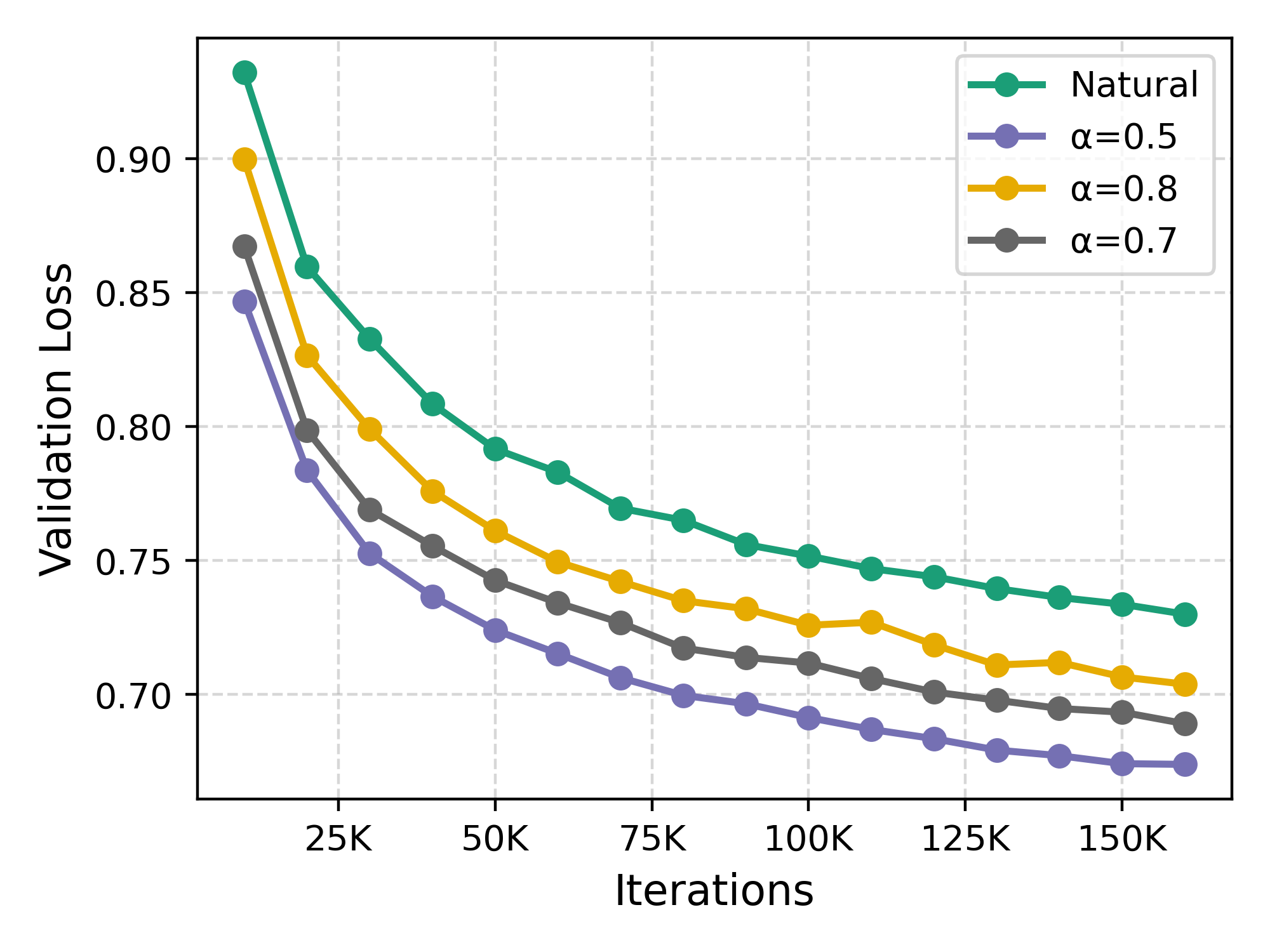}
    \caption{The balanced sampler improves the validation loss}
    \label{fig:balanced_sampler_results}
\end{figure}

Inspired by the SALMONN architecture~\cite{tangsalmonn}, we opt for a two-layer window-level Q-former projector and temporal downsampler as the speech modality adapter. The Q-former architecture~\cite{li2023blip} was introduced to convert an encoded image into a small number of textual tokens that will be consumed by an LLM. The idea is to have a fixed number of trainable queries that can attend to each other as well as to the image embeddings. The authors in~\cite{tangsalmonn} extended the application of Q-former to variable-length sequences as follows. Given $N$ trainable queries $\mathbf{Q}=\mathbf{q}_1\ldots\mathbf{q}_N$ and $\mathbf{X}=\mathbf{x}_1\ldots\mathbf{x}_T$ an acoustic embedding sequence of length $T$ computed by the acoustic encoder, let $K\ge N$ denote a block (or window) size such that $K~\rm{mod}~N=0$. $\mathbf{X}$ is converted to $\mathbf{Y}=\mathbf{y}_1\ldots\mathbf{y}_{N*\left\lceil T/K\right\rceil}$ by

\begin{eqnarray*}
 \mathbf{y}_{(i-1)*N+1}\ldots\mathbf{y}_{i*N}=\text{Q-former}(\mathbf{Q},\mathbf{x}_{(i-1)*K+1}\ldots\mathbf{x}_{i*K}),\\
i=1\ldots \left\lceil T/K\right\rceil      
\end{eqnarray*}
~~\\
where it is understood that $\mathbf{X}$ is padded with zero vectors from $T+1\ldots K*\left\lceil T/K\right\rceil$. Note that the Q-former performs a temporal downsampling of the acoustic embeddings by a factor of $K/N$. The Q-former is trained jointly with LoRA adapters of rank 64 applied to the query and value projection matrices for all the attention layers of the Granite LLMs. The CTC speech encoder is kept frozen during this training phase. The training criterion is the next token prediction cross-entropy loss applied to the target ASR or AST transcripts. The training was performed with AdamW minibatch SGD over three epochs, 660000 updates, a peak learning rate of 1e-4 with a warm-up phase of 1000 steps, and a batch size of 128 utterances distributed over 32 H100 GPUs. Additionally, we used balanced sampling to ensure that we get an adequate representation for corpora with fewer samples. Specifically, if we have $L$ corpora with number of samples $N_1\ldots N_L$ and a factor $\alpha\in[0,1]$, we sample from corpus $i$ with probability $\frac{N_i^\alpha}{\sum_{j=1}^L N_j^\alpha}$. The extreme cases are $\alpha=0$ where we sample from each corpus with uniform probability and $\alpha=1$ which corresponds to the natural distribution. In practice, $\alpha=0.6$ achieved good performance across a majority of corpora. Figure~\ref{fig:balanced_sampler} illustrates how the balanced sampler flattens the data distribution and in Figure~\ref{fig:balanced_sampler_results} we show how lower $\alpha$ values lead to better validation losses.

\begin{figure}[htb]
    \centering
    \includegraphics[width=0.98\linewidth]{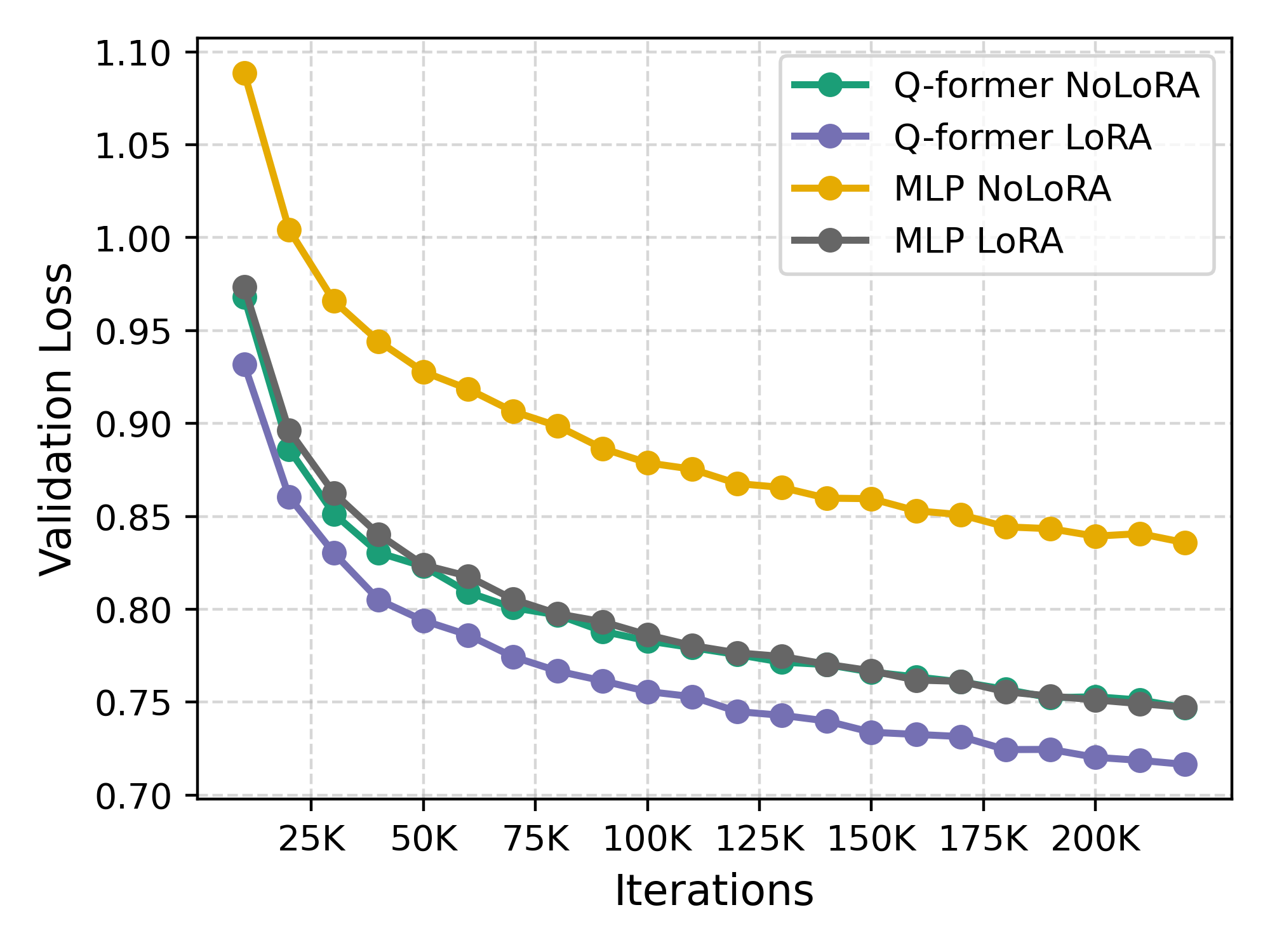}
    \caption{Validation losses for Q-former and MLP projectors, with and without applying LoRA to the LLM weights}
    \label{fig:linear_qformer_lora_nolora}
\end{figure}

In Figure~\ref{fig:linear_qformer_lora_nolora} we compare the validation losses of a Q-former projector and a 2-layered MLP projector, with or without LoRA adapters with rank 64 applied to the query and value projection matrices of attention blocks in the LLM. 
In Table~\ref{proj-wer}, we compare Q-former projectors with different block sizes $K$ and number of queries $N=K/5$ against a 2-layer MLP projector similar to~\cite{ma2024embarrassingly,abouelenin2025phi} and also a projector that uses cross-attention from temporally-downsampled acoustic embeddings (queries) to the LLM text embedding table (keys and values). All experiments use a temporal downsampling factor of 5 for acoustic embeddings, granite-3.1-8b-instruct as the base LLM and include the GigaSpeech, SPGI and TED\_LIUM datasets in the training data (but exclude YODAS).

\begin{table}[htb]
\begin{center}
\begin{tabular}{|l|c|c|c|c|c|c|c|c|c|}\hline
Projector       &  CV  & GS & MLS & LSc & LSo & SPGI & AMIi & AMIs & Vox\\ \hline
Qf $K=25$ &  9.7 & 10.2 &4.9 &1.4 &3.1  &2.1   & 9.7  &27.5  &6.5\\ 
Qf $K=15$ &  9.6 & 10.1 &4.9 &1.4 &3.0  &2.1   & 10.4 &27.5  &6.5\\
Qf $K=10$ &  9.6 & 10.1 &4.8 &1.4 &3.2  &2.1   & 10.4 &27.8  &6.4\\ \hline
MLP       & 10.4 & 10.2 &5.0 &1.4 &3.3  &2.2   & 10.0 &28.1  &6.6\\ \hline
x-attn    & 10.4 & 10.6 &5.2 &1.5 & 3.4 & 2.4  & 10.8 & 27.7 & 7.2\\ \hline
\end{tabular}
\end{center}
\caption{\label{proj-wer}ASR performance of window Q-former projector with different block sizes compared to projectors using MLP and cross-attention to LLM text embeddings table (granite-3.1-8b-instruct was used as the base LLM).}
\end{table}

We observe that Q-former outperforms both MLP and cross-attention to LLM text embeddings table projectors and a block size of $K=15$ frames and $N=3$ queries strikes a good balance between computational complexity and ASR performance. With these settings, the original 100 Hz logmel frame rate is reduced to a 10 Hz acoustic embeddings rate coming out of the Q-former into the LLM (2x at the input of the CTC encoder and $K/N=5$x after Q-former). 

In the next series of experiments, we train on Apache 2.0 compatible corpora only (no GigaSpeech, SPGI and TED\_LIUM but include YODAS) and look at the effect of the text LLM on ASR performance of the speech-aware LLM. In particular, in Table~\ref{llm-wer} we compare the use of granite-3.2-8b-instruct, granite-3.3-8b-instruct and granite-3.3-2b-instruct as the base LLMs. The results were obtained with batched inference with 4 samples per batch, beam search with a beam of 4, and a token repetition penalty of 3.0 applied only to generated tokens~\cite{keskar2019ctrl}. 

\begin{table}[htb]
\begin{center}
\begin{tabular}{|l|c|c|c|c|c|c|c|c|c|}\hline
LLM            &  CV  & GS   & MLS & LSc & LSo & SPGI & AMIi & AMIs & Vox \\ \hline
granite-3.2-8b & 8.0  & 10.5 & 4.8 & 1.5 & 3.1 & 3.0  & 9.2  & 26.0 & 5.9 \\ \hline
granite-3.3-8b & 7.0  & 10.5 & 4.7 & 1.5 & 3.0 & 3.2  & 9.2  & 26.1 & 5.8 \\ \hline
granite-3.3-2b & 8.1  & 10.8 & 5.2 & 1.6 & 3.4 & 3.6  & 9.4  & 26.7 & 6.2 \\ \hline
\end{tabular}
\end{center}
\caption{\label{llm-wer}ASR performance as a function of LLM choice for models trained on Apache 2.0 compatible corpora only.}
\end{table}

We note that models trained with granite-3.2-8b-instruct and granite-3.3-8b-instruct exhibit comparable performance (except for CommonVoice) and that, unsurprisingly, the 8B parameter variants outperform the 2B parameter model across all corpora.

\begin{figure*}[htb]
    \hfill
    \centering
    \centerline{\includegraphics[width=0.85\textwidth]{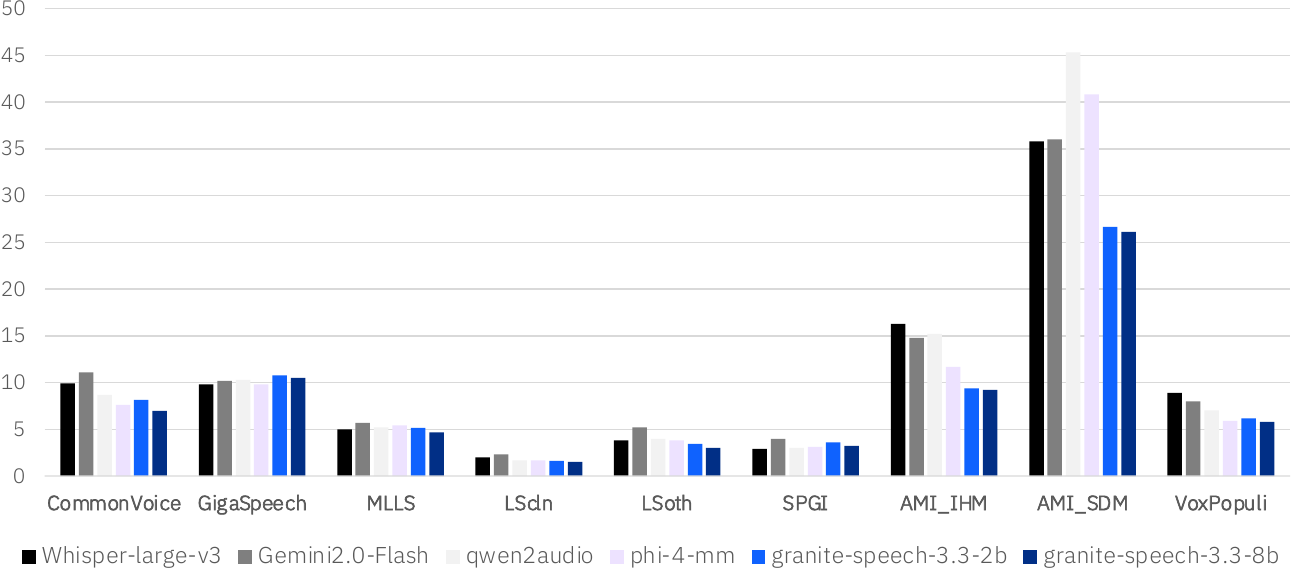}}
\caption{\label{wer-comp} Word error rate comparison between 2B and 8B parameter granite-speech-3.3 models and leading SLMs on public benchmarks for English ASR.}
\end{figure*}

In Figure~\ref{wer-comp} we compare granite-speech-3.3-2b and granite-speech-3.3-8b (last two rows of Table~\ref{llm-wer}) against other leading SLMs in the category of less than 8B parameters as well as dedicated ASR systems such as OpenAI's Whisper large v3. We remark that our 8B parameter model achieves the lowest WERs on all corpora except GigaSpeech and ties on SPGI probably because both GigaSpeech and SPGI train splits were excluded from the training data. Moreover, the 2B model achieves competitive performance, especially on AMI where it comes very close to the 8B model. AMI is a difficult corpus of distant microphone meeting recordings, suggesting that the smaller model may be more robust to challenging acoustic environments. 
 
\subsection{Speech translation}
\label{ssec:ast}

CoVoST2 is the most widely used speech translation dataset~\cite{wang2021covost}, but its size is small in scale compared to typical speech recognition training corpora. Its license is also more restrictive in terms of commercial use.
Synthetic speech translation data can be generated in two ways: by applying speech synthesis to machine translation datasets to produce source-language audio, or by translating transcriptions from speech recognition datasets to obtain target-language text. We adopt the latter approach.

Using Phi-4~\cite{abdin2024phi}, an LLM with excellent machine translation performance, we translated text from CoVoST2 test data and found that the BLEU score for en$\to$de was 34.3 and en$\to$ja was 29.2, which is not high enough for good quality training data generation. Instead, we select data based on the assumption that, if the output of two different models for a text in a source language is close or identical, then the translation is likely to be more reliable. Concretely, we input the same source language text into two machine translation models and calculate the similarity of the output of the two translation results. WER, BLEU, and cosine distance were used as similarity measures, and CoVoST2 test data was used to investigate what threshold values of the measures can effectively extract reliable translation results.

\begin{figure}[htb]
    \centering
    \includegraphics[width=1.00\linewidth]{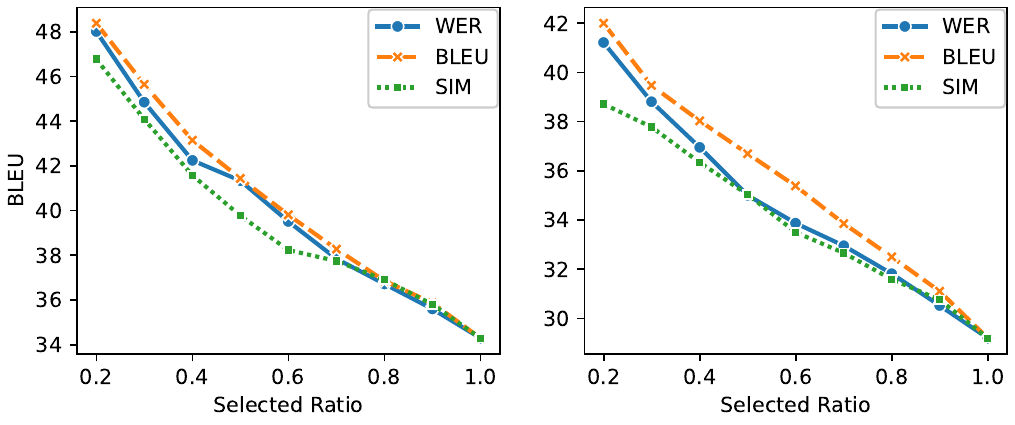}
    \caption{Results of ensemble filtering for en$\to$de (left) and en$\to$ja (right), where the x-axis is the percentage of data selected by the threshold and the y-axis is the average BLEU score of the selected data.}
    \label{fig:ast_data_filtering_bymetrics}
\end{figure}

\begin{figure}[htb]
    \centering
    \includegraphics[width=1.00\linewidth]{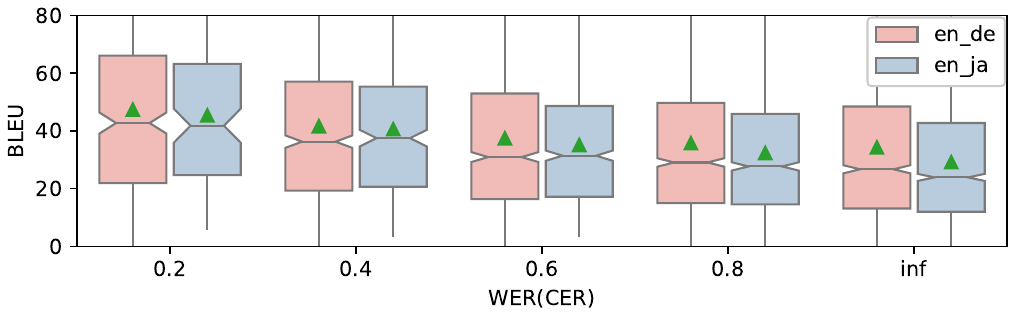}
    \caption{Distribution of BLEU scores for subsets with WER (CER for ja) as selection metric (triangles indicate averages).}
    \label{fig:ast_wer_distribution}
\end{figure}

Figure~\ref{fig:ast_data_filtering_bymetrics} shows the amount of data in the selected subset and the average BLEU score for that subset depending on which similarity index threshold was applied. Phi-4 was used as the primary translation model and Granite-3.2 as the secondary model, because the BLEU score of CoVoST2 using granite-3.2 was 29.9 for en$\to$de, 21.9 for en$\to$ja, and phi-4 had better translation performance. In the case of WER and BLEU, the output of the main model was computed as a reference and the sub-model as a hypothesis. The cosine distance was calculated using the distance between the output vectors obtained by inputting the outputs into the multilingual-sentence-transformer~\cite{reimers2020making}. When comparing WER and BLEU as selection thresholds, we observe that the trend is almost the same, and reducing the fraction of selected data to 0.2 improves the average BLEU score by more than 10 points. The cosine distance was less effective as a selection metric compared to WER and BLEU. Figure~\ref{fig:ast_wer_distribution} shows the distribution of BLEU scores for the subset data when WER is used as the selection metric, indicating that, the lower the WER, the better quality machine translation results with higher BLEU scores are extracted. We also examined the average length of the subsets to ensure that the selected subset data were not biased toward short sentences and found that the average length was consistently almost the same for all subsets.

\begin{figure*}[htb]
    \hfill
    \centering
    \centerline{\includegraphics[width=0.85\textwidth]{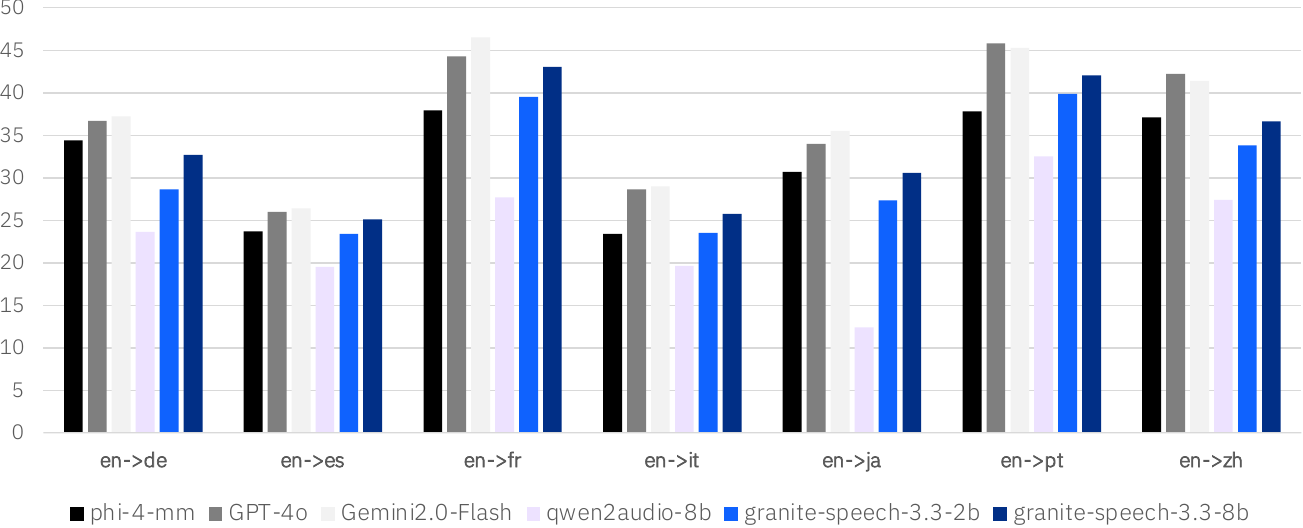}}
\caption{\label{bleu2-comp} BLEU scores comparison between 2B and 8B parameter granite-speech-3.3 models and leading SLMs on FLEURS En-X speech translation.}
\end{figure*}

Based on these experiments, we compared several translation models and finally generated training data using Phi-4 as the primary translation model and MADLAD-3B/10B~\cite{kudugunta2023madlad} as the secondary translation model for threshold calculation. The CommonVoice English training data was translated using these two models, with WER=0.3 for en$\to$de translation and CER=0.4 for en$\to$ja, and the Phi-4 translations were used as training data. After filtering, the amount of retained data is less than half of the original CommonVoice data, but the translations are likely to be reliable.

\begin{figure}[htb]
    \centering
    \includegraphics[width=1.00\linewidth]{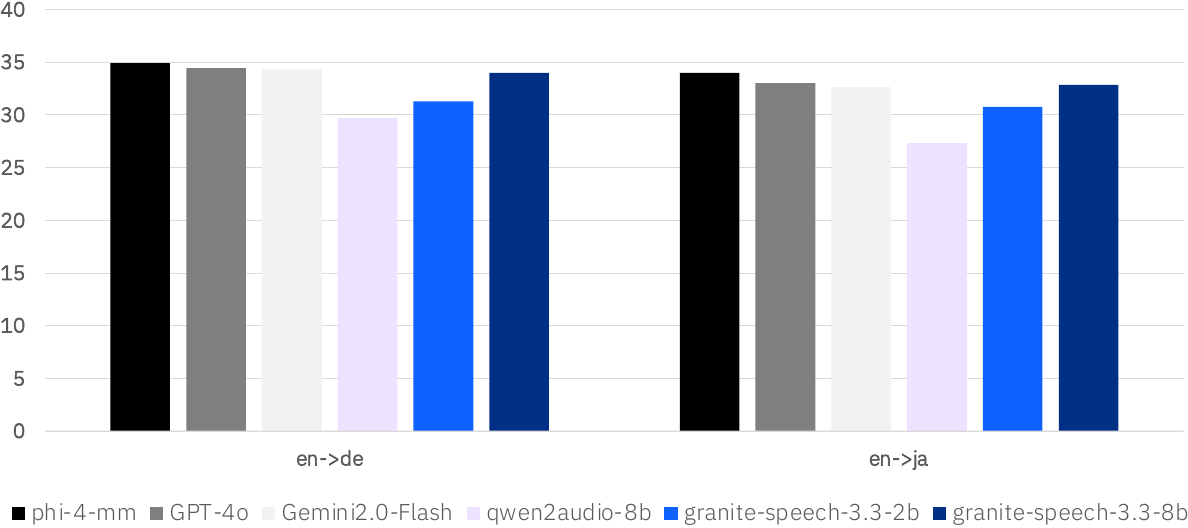}
\caption{\label{bleu1-comp} BLEU scores comparison between 2B and 8B parameter granite-speech-3.3 models and leading SLMs on CoVost2 En-De, En-Ja speech translation.}
\end{figure}

In Figures~\ref{bleu2-comp} and~\ref{bleu1-comp} we show the speech translation performance for the granite-speech-3.3-2b and granite-speech-3.3-8b models from Table~\ref{llm-wer} in comparison to other leading SLMs on FLEURS~\cite{conneau2023fleurs} and CoVost2, respectively. While our models trail the leading SLMs on FLEURS, they achieve competitive performance on CoVost2 En-De and En-Ja. It is also worth mentioning that our 8B model has a noticeably better translation performance than the 2B variant for both corpora and across all conditions.

\subsection{Safety}
\label{sec:safety}
Our safety assessment of the Granite-speech LLM employed a rigorous, multi-stage protocol to ascertain whether coupling audio clips with harmful textual instructions, sourced from established safety benchmarks (BOLD~\cite{bold_2021}, AttaQ~\cite{attaq_2023}, Toxigen~\cite{toxigen_2022}), could compromise the model’s instruction-aligned behavior. In the first stage, each toxic instruction was paired with low-amplitude noise segments; the system invariably repeated the prompt verbatim without executing or expanding upon its content. In the second stage, we presented toxic instructions together with content-rich audio excerpts from established corpora. The model repeated each instruction verbatim, then accurately transcribed the audio without carrying out the harmful directive. These results show that the speech interface maintains the strong refusal behavior of the underlying text model, preventing unsafe responses even when faced with noisy or complex audio inputs.

\section{Conclusion}
\label{sec:conclusion}
In this paper, we have described the design choices and experimental setups for a class of speech-aware LLMs focused on English ASR and English-to-foreign speech translation. On the acoustic encoder side, we opted for conformer CTC with character-level tokenization, block self-attention and self-conditioned CTC. On the speech modality adapter side, we showed the benefit of using window-level Q-former over MLP and cross-attention from the acoustic embeddings to the LLM embeddings table. Additionally, we discussed multi-task prompt formulation, chain-of-thought speech translation, and data generation and selection for AST. Importantly, we finished by addressing safety considerations of the proposed Granite-speech LLMs. 

Future work will primarily address the biggest gap in the current models which is multilingual ASR. We also plan to produce richer ASR transcripts by providing time marks and speaker turn information. In a connected line of research, we intend to look at context-sensitive ASR using contextual biasing for keyword recognition or previous dialog turns and TTS synthesis of text dialogs used for instruction fine-tuning Granite LLMs for specific tasks. Last but not least, we plan to incorporate paralinguistic information like speaker emotion into our model to enhance the overall end-to-end user interaction experience.

\bibliographystyle{IEEEtran}
\bibliography{main}

\end{document}